# Many body renormalization of the minimal conductivity in graphene

F. Guinea[a]
M. I. Katsnelson[b]

[a]Instituto de Ciencia de Materiales de Madrid. CSIC. Sor Juana Ines de la Cruz 3. 28049 Madrid. Spain
[b]Radboud University Nijmegen, Institute for Molecules and Materials, Heyndaalseweg 135, NL-6525AJ Nijmegen, The Netherlands

Abstract

The conductance of ballistic graphene at the neutrality point is due to coherent electron tunneling between the leads, the so called pseudodiffusive regime. The conductance scales as function of the sample dimensions in the same way as in a diffusive metal, despite the difference in the physical mechanisms involved. The electron-electron interaction modifies this regime, and plays a role similar to that of the environment in macroscopic quantum phenomena. We show that interactions, and the presence of external gates, change substantially the transport properties, and can lead to a diverging resistivity at the neutrality point.

The electrical conductivity of solids is determined by electron or hole excitations at the Fermi level. One of the most striking features of graphene is its finite metallic conductivity when the Fermi surface shrinks to a point, and the density of charge carriers vanishes[1,2]. The origin of this minimal conductivity is a problem of fundamental relevance. Early experiments suggested that the conductivity at the neutrality point was of order of a conductance quantum, while recent measurements in high mobility samples give a much lower value[3,4]. Carriers become localized when the conductivity drops below the quantum unit, but in graphene localization is suppressed by "Klein" tunneling[5]. Calculations show that graphene remains metallic at the neutrality point. The same conclusion can be reached assuming that graphene is defect-free and ballistic at the neutrality point, due to an essentially quantum phenomenon, transmission via evanescent waves[6-8]. We analyze here the effect of the electron-electron interaction in this regime, and, thus, on the minimal conductivity of graphene.

Experiments show that the Coulomb interaction between electrons change substantially the electronic properties near the Dirac point in high mobility suspended systems[9]. The effect of interactions on the conductivity of graphene at the Dirac point has been addressed theoretically, using diagrammatic methods and starting from the Kubo expression for the conductivity[10-13]. The conclusion of these works is that the metallic nature of graphene near the Dirac point is not changed by interactions. We consider here the alternative description where the conductance of a ballistic graphene sample is studied using Landauer's formalism, adding later the electron-electron interaction, and come to essentially different conclusions. It turns out that the interaction effects suppress essentially the transport via evanescent waves leading to temperature (or sample-size) dependent minimal conductivity, in agreement with recent experimental observations[4]. Conduction in a perfect ballistic graphene sample at the Dirac point is due to tunneling of electrons with well defined momentum parallel to the direction of current[6,7]. The summation of the transmission coefficients of all these parallel momentum channels give rise to a conductance inversely proportional to the system length, defined as the transport direction, and inversely proportional to the perpendicular direction. This scaling with the sample dimensions is the same as in a diffusive metal, leading to the term "pseudodiffusive regime"[7]. This approach can be generalized to graphene bilayers[14-16], and to samples of arbitrary shapes[17,18]. We assume that the tunneling electrons can excite electron-hole pairs and other electronic excitations of the system, which are considered to be independent degrees of freedom. This approach can be justified by replacing the excitations of the electronic system by bosons, each of which is weakly coupled to the tunneling electron[19,20]. This approach has proven very useful in the study of quantum tunneling of particles interacting with their environment. Formally, the method can be viewed as a resummation of bubble diagrams similar to the Random Phase Approximation[21].

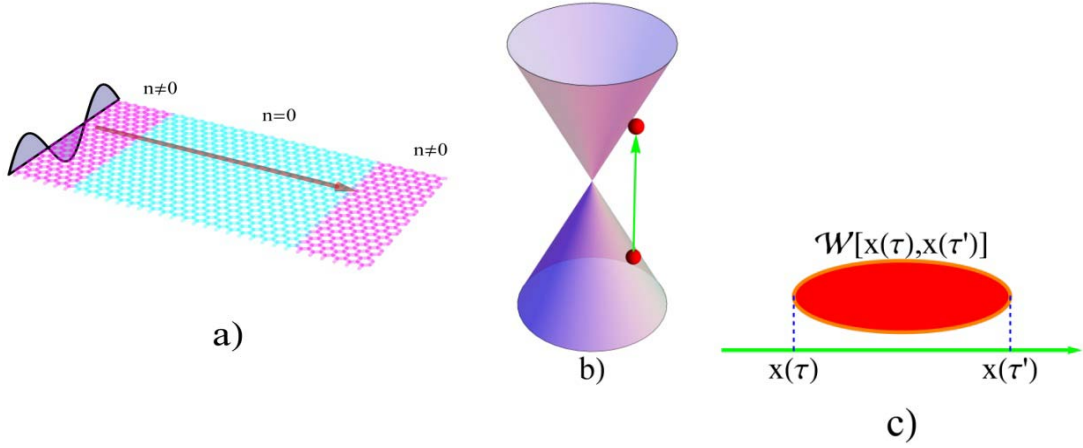

FIG. 1. (a) Sketch of the processes considered in the text. An electron wavepacket, coherent in the direction normal to the direction of the current, is transferred between two electrodes. b) The tunneling process is accompanied by the emission of electron-hole pairs. c) Lowest order diagram which describes the process.

We analyze transport through a rectangular graphene sample of dimensions $L_x, L_y$, where the $x$ axis is the current direction, and $L_x, L_y \gg a$, where $a$ is the lattice spacing, see Fig. 1. For simplicity we use periodic boundary conditions along the $y$ direction[6]. We assume that the wave function is coherent along the $y$ direction and the transverse momentum is quantized, $k_y = 2\pi n/L_y$, where $n$ is an integer. The tunneling along the $x$ direction is studied by estimating the optimal path in imaginary time, and adding to the action along that path the corrections due to the interactions with the environment. The barrier through which tunneling takes place is $V(x) = \hbar v_F k_y$ for $0 \leq x \leq L_x$, and $v_F$ is the Fermi velocity. The path under the barrier is simply $x(\tau) = v_F \tau$, with $0 \leq \tau \leq L_x/v_F$.

The tunneling amplitude, in the absence of interaction effects, is $T_0(k_y) \cong e^{-k_y L_x}$. The correction to the action due to the interactions with the environment can be written as[22]

$$\delta S = \frac{1}{2} \int_{-\infty}^{+\infty} d\tau \int_0^\beta d\tau' \int_{-\infty}^{+\infty} \frac{dq}{2\pi} e^{iq[x(\tau)-x(\tau')]} v_q^2 \langle \hat{T}[\rho_q(\tau)\rho_{-q}(\tau')]\rangle \quad (1)$$

where $\beta = 1/T$ is the inverse temperature, $v_q$ is the Fourier component of the Coulomb interaction, $\rho_q$ is the electron density operator and $\hat{T}$ is the time-ordering operator. Using the fluctuation-dissipation theorem and proceeding further as in[21] we come to the expression

$$\delta S = \frac{1}{2}\int_{-\infty}^{+\infty} d\tau \int_0^\beta d\tau' \int_{-\infty}^{+\infty}\frac{dq}{2\pi}\int_{-\infty}^{+\infty} d\omega\, e^{iq[x(\tau)-x(\tau')-\omega|\tau-\tau'|]}W(q,\omega) \quad (2)$$

where $W(q,\omega)$ is a density of states which includes the density of states of modes in the environment, and their coupling to the tunneling electron

$$W(q,\omega) = \frac{v_q}{\epsilon(q,\omega)} \quad (3)$$

and $\epsilon(q,\omega)$ is the dielectric function. The tunneling amplitude is finally. $T(k_y) \cong T_0(k_y)e^{-\delta S}$.

For the effective one-dimensional (1D) problem defined here, we have[22]

$$v_q \simeq \begin{cases} -\frac{2e^2}{\epsilon_0}\log(qL_y) & qL_y \ll 1 \\ \frac{2\pi e^2}{\epsilon_0 qL_y} & 1 \ll qL_y \end{cases} \quad (4)$$

where $\epsilon_0$ is the dielectric constant of the environment. This expression interpolates between the expected 1D behavior for $qL_y \ll 1$, and the 2D Coulomb interaction, normalized to the width of the sample for $qL_y \gg 1$.

We consider first an environment made up of the electron-hole excitations of graphene at the neutrality point. The dielectric function can be written as $\epsilon(q,\omega) = 1 + v_q \chi_{1D}(q,\omega)$ where $\chi_{1D}(q,\omega)$ is the polarization function of our 1D problem. We assume that $L_y \leq L_x$. Then, the leading contributions to $\delta S$ come from $q \simeq L_x^{-1}$, and we use the lower line in eq.(4). The dielectric function of a graphene ribbon was calculated in[24]. In wide ribbons, $L_y \gg a$, the Coulomb potential does not mix subbands, and we can approximate[23]

$$\chi_{1D}(q,\omega) \approx L_y \chi_{2D}(q,\omega) \approx L_y \frac{q^2}{4\sqrt{v_F^2 q^2 - \omega^2}} \quad (5)$$

In the ballistic regime, where $x(\tau) = v_F \tau$, the time integrals in eq. 2 can be reduced to:

$$\delta S_G \approx \frac{L_x}{(2\pi)^2 L_y v_F}\int_{L_x^{-1}}^{a^{-1}} dq \int_{v_F q}^{v_F a^{-1}} d\omega \int_0^{L_x/v_F} d\tau \frac{4(2\pi e^2/\epsilon_0)^2}{16(\omega^2 - v_F^2 q^2)^2 + (2\pi e^2/\epsilon_0)^2 q^2} e^{iqv_F\tau} e^{-\omega\tau} \approx$$

$$\frac{L_x}{(2\pi)^2 L_y v_F}\int_{L_x^{-1}}^{a^{-1}} dq \int_{v_F q}^{v_F a^{-1}} d\omega \frac{4(2\pi e^2/\epsilon_0)^2}{16(\omega^2 - v_F^2 q^2)^2 + (2\pi e^2/\epsilon_0)^2 q^2}\frac{\omega}{\omega^2 + v_F^2 q^2} \approx \frac{L_x}{8\pi L_y}\frac{\alpha^2}{4\sqrt{2}+\alpha}\log\left(\frac{L_x}{a}\right) \quad (6)$$

where $\alpha = (2\pi e^2)/(\epsilon_0 v_F)$, and we have kept only the leading term in $L_x/a$.

We now consider the changes induced in the environment by the presence of a metallic layer. We describe the metal in terms of its density of states, $v_{1D} \approx L_y v_{2D}$, Fermi velocity, $v_F^M$, Fermi energy, $\epsilon_F$, Fermi momentum, $k_F$, mean free path, $\ell$, and diffusion coefficient, $D = (v_F^M \ell)/2D$. The polarizability of the metal, for $\omega \leq \epsilon_F$ and $q \leq k_F$, can be approximated by

$$\chi_{1D}^M(q,\omega) \approx \begin{cases} \frac{v_{1D} D q^2}{i\omega + D q^2} & q \leq \ell^{-1} \\ \frac{v_{1D} v_F^M q}{i\omega + v_F^M q} & \ell^{-1} \leq q \leq k_F \end{cases} \quad (7)$$

In the RPA approximation, the retarded interaction is given, approximately, by $W(q,\omega) \approx [\chi_{1D}^M(q,\omega)]^{-1}$. Then, the value of $\delta S_M$ can be divided into a diffusive and a ballistic contribution

$$\delta S_M = \delta S_d + \delta S_b \approx \frac{L_x^2}{4\pi g \ell L_y} + \frac{L_x}{8\pi L_y} \log(g) \quad (8)$$

where $g = k_F \ell$ is the conductivity of the metallic layer. At finite temperatures, $T \neq 0$, the momentum cutoff becomes $q_c \approx \text{Max}(L_x^{-1}, T/v_F)$. Fig. 2 shows the temperature dependence of the resistivity taking into account a neutral graphene environment, $\delta S_G$, in eq. 6, and a metallic environment, $\delta S_M$ in eq. 8. The parameters used for the metallic layer are appropriate for graphene away from the neutrality point, see Fig. 2. For this choice of parameters, the final conductivity is determined by the contribution from the diffusive modes of the metal.

The pseudodiffusive regime can be generalized to situations with external magnetic fields[24]. The presence of a magnetic field changes the conductivity in the metal, due to the suppression of coherence effects. In addition, the classical trajectories in the neutral ballistic graphene layer are modified on scales comparable to the magnetic length, $\ell_B$. A simple perturbative estimate of the self energy in the presence of a magnetic field shows that the effective interaction is modified[22]. $W(q,\omega) \approx \int dq' e^{-(q-q')^2/\ell_B^2} W(q',\omega)$. This $B$ dependent broadening suggests the use of the lower cutoff, $q_c \approx \text{Max}(L_x^{-1}, T/v_F, \ell_B^{-1})$. The magnetic field dependence of the inverse conductance using this approximation is shown in Fig. 3. Note that a numerical constant $c$ in the definition of $q_c$ will change the temperature and magnetic field scales, although not the qualitative trends.

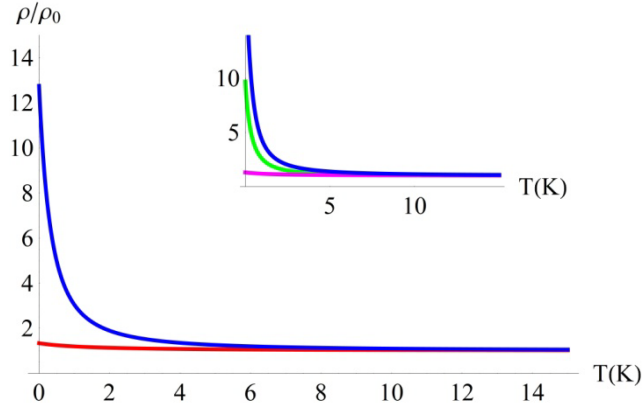

FIG. 2. Temperature dependence of the inverse conductance, normalized to the non interaction value, $\sigma_0 = e^2/(\pi\hbar)$, for $L_x = 4\mu$, $L_y = 1\mu$. Red: Contribution from the graphene excitations, $\delta S_G$, eq. 6. Blue: Contribution from a metallic layer, $\delta S_M$, eq. 8. The two terms which describe the contribution from the metal, $\delta S_d$ and $\delta S_b$ are shown in the inset. Green: diffusive part, $\delta S_d$ in eq. 8. Magenta: ballistic part, $\delta S_b$ in eq. 8. The carrier density in the metal is $n = 10^{11} \text{cm}^{-1}$, and the elastic mean free path is $\ell = 100$nm.

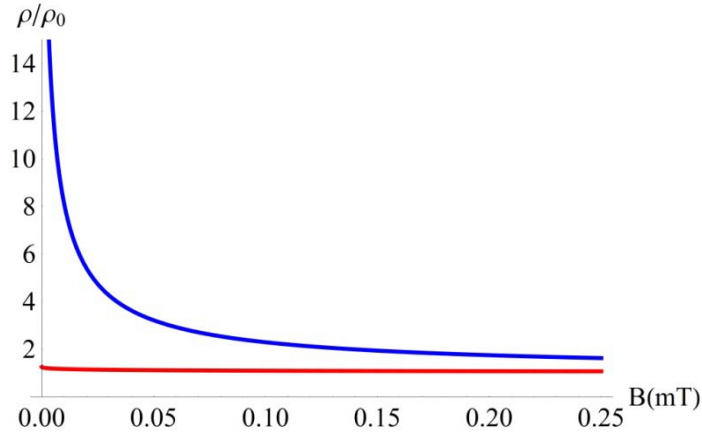

FIG. 3. Magnetic field dependence of the inverse conductance for $T = 1$K. The remaining parameters are the same as in Fig. 2.

Another situation where electron tunneling is relevant is ballistic transport through a p-n junction[25,26]. The properties of a planar p-n junction are determined by the electric _field $\mathcal{E}$ when the potential lies close to the Dirac energy, $V(x) \approx e\mathcal{E}x$. Electrons with a well defined parallel momentum, $k_y$, and a dispersion $\varepsilon_k = v_F\sqrt{k_x^2 + k_y^2}$ have a gap of forbidden energies $\Delta_{k_y} = v_F k_y$. Hence, an electron with momentum $k_y$ must tunnel through a barrier through the region $-(v_F k_y)/(2\mathcal{E}) \leq x \leq (v_F k_y)/(2\mathcal{E})$. The probability of tunneling is[8,25]

$T_0(k_y) \approx e^{-(v_F k_y^2)/\mathcal{E}}$. Interactions suppress tunneling through p-n junctions in the manner discussed above, with the replacement $L_x \leftrightarrow (v_F k_y)/\mathcal{E}$ in eq. 6 and eq. 8. For example, instead of eq. 6 we have:

$$T(k_y) \approx e^{-\frac{v_F k_y^2}{\mathcal{E}}} \left(\frac{v_F |k_y|}{a\mathcal{E}}\right)^{-\frac{v_F |k_y|}{8\pi L_y \mathcal{E} 4\sqrt{2}+\alpha} \frac{\alpha^2}{}} \qquad (9)$$

This renormalization changes essentially the angular dependence of the tunneling probability for very small angles, $|k_y| \ll \alpha^2/(8\pi L_y)$. At the same time, $T = 1$, an exact property for normal incidence[8], remains unchanged when the electron-electron interactions taken into account. The changes induced in the angular dependence of the transmission are shown in Fig. 4.

The dependence of the total conductance, $\propto L_y/(2\pi) \int dk_y\, T(k_y)$, on the electric field is changed, due to the renormalization in eq. (10), from $\sqrt{\mathcal{E}}$ (which corresponds to the Schwinger effect, with the pair intensity production $P \propto \mathcal{E} G \propto \mathcal{E}^{3/2}$, see[27,28] and references therein) to $P \propto \mathcal{E}^2$. The crossover takes place at the electric field $\mathcal{E} \approx (\alpha^2 v_F)/(8\pi L_y^2)$.

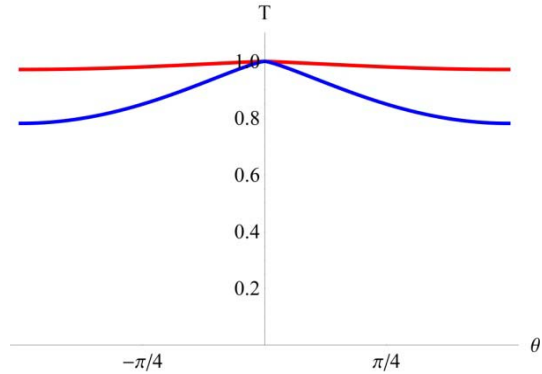

FIG. 4. Angular dependence of the correction due to interactions in a p-n junction. The p-n junction has a length of 1 nm and separates two regions of densities $n = \pm 10^{12} \text{cm}^{-2}$. The value of $\alpha$ is 2.2/2. Red: Intrinsic effect in graphene at the neutrality point, $\delta S_G$, eq. 6. Blue: Effect of a metallic layer, $\delta S_M$, eq. 8. The charge density in the metal is $n = 10^{13}$ cm$^{-2}$.

Tunneling between localized states is another mechanism which gives rise to a finite conductivity of graphene at the neutrality point[29]. The interaction effects discussed here will also influence this mechanism[22]. The lack of an intrinsic limit to the conductivity of ballistic graphene at the neutrality point suggests new ways to manipulate its value. The combination of quantum tunneling and interactions implies that ballistic graphene at the neutrality point can be used to study dephasing processes under a variety of external probes.

F.G. acknowledges financial support from MINECO, Spain, through grant FIS2011-23713, and the European Research Council Advanced Grants program, through grant 290846. MIK acknowledges a financial support from FOM, The Netherlands.

# References

[1] K. S. Novoselov, A. K. Geim, S. V. Morozov, D. Jiang, Y. Zhang, S. V. Dubonos, I. V. Grigorieva, and A. A. Firsov, Science, **306**, 666 (2004).

[2] K. S. Novoselov, D. Jiang, F. Schedin, T. J. Booth, V. V. Khotkevich, S. V. Morozov, and A. K. Geim, Proc. Natl. Acad. Sci. U.S.A., **102**, 10451 (2005).

[3] L. A. Ponomarenko, A. K. Geim, A. A. Zhukov, R. Jalil, S. V. Morozov, K. S. Novoselov, V. V. Cheianov, V. I. Fal'ko, K. Watanabe, T. Taniguchi, and R. V. Gorbachev, Nature Phys. **7**, 958 (2011).

[4] F. Amet, J. R. Williams, K. Watanabe, T. Taniguchi, and D. Goldhaber-Gordon, (2012), arXiv:1209.6364.

[5] M. I. Katsnelson, K. S. Novoselov, and A. K. Geim, Nature Phys., **2**, 620 (2006).

[6] M. I. Katsnelson, Eur. J. Phys. B, **51**, 157 (2006).

[7] J. Tworzydo, B. Trauzettel, M. Titov, A. Rycerz, and C. W. J. Beenakker, Phys. Rev. Lett., **96**, 246802 (2006).

[8] M. I. Katsnelson, Graphene: Carbon in Two Dimensions (Cambridge University Press, Cambridge, 2012).

[9] D. C. Elias, R. V. Gorbachev, A. S. Mayorov, S. V. Morozov, A. A. Zhukov, P. Blake, K. S. Novoselov, A. K. Geim, and F. Guinea, Nature Phys., **7**, 701 (2011).

[10] L. Fritz, J. Schmalian, M. Müller, and S. Sachdev, Phys. Rev. B, **78**, 085416 (2008).

[11] V. Juricic, O. Vafek, and I. F. Herbut, Phys. Rev. B, **82**, 235402 (2010).

[12] I. V. Gornyi, V. Yu. Kachorovskii, A. D. Mirlin, Phys. Rev. B **86**, 165413 (2012).

[13] A. Giuliani and V. Mastropietro, Phys. Rev. B, **85**, 045420 (2012).

[14] M. I. Katsnelson, Eur. Phys. J., **52**, 151 (2006).

[15] I. Snyman and C. W. J. Beenakker, Phys. Rev. B, **75**, 045322 (2007).

[16] J. Cserti, A. Csordas, and G. Dvid, Phys. Rev. Lett., **99**, 066802 (2007).

[17] M. I. Katsnelson and F. Guinea, Phys. Rev. B, **78**, 075417 (2008).
[18] A. Rycerz, P. Recher, and M. Wimmer, Phys. Rev. B, **80**, 125417 (2009).
[19] A. O. Caldeira and A. J. Leggett, Phys. Rev. Lett., **46**, 211 (1981).
[20] A. O. Caldeira and A. J. Leggett, Ann. Phys. (N.Y.), **149**, 374 (1983).
[21] F. Guinea, Phys. Rev. Lett., **53**, 1268 (1984).
[22] See Supplementary Information.
[23] L. Brey and H. A. Fertig, Phys. Rev. B, **75**, 125434 (2007).
[24] E. Prada, P. San-Jose, B. Wunsch, and F. Guinea, Phys. Rev. B, **75**, 113407 (2007).
[25] V. V. Cheianov and V. I. Fal'ko, Phys. Rev. B, **74**, 041403 (2006).
[26] L. M. Zhang and M. M. Fogler, Phys. Rev. Lett., **100**, 116804 (2008).
[27] J. C. Kao, M. Lewkowicz, and B. Rosenstein, Phys. Rev. B, **82**, 035406 (2010).
[28] M. I. Katsnelson, G. E. Volovik, and M. A. Zubkov, Ann. Phys. (N. Y.), **331**, 160 (2013).
[29] M. Titov, Europhys. Lett. **79**, 17004 (2007).

**Suplementary information**

*-Effective potential, equation (4).*
The Coulomb potential between coherent electron waves localized at positions $x$ and $x'$ is defined as

$$V(x - x') = \frac{1}{L_y^2} \int_0^{L_y} dy \int_0^{L_y} dy' \frac{e^2}{\epsilon_0 \sqrt{(x-x')^2 + (y-y')^2}} \quad (S1)$$

and

$$v_q = \frac{1}{2\pi} \int_{-\infty}^{\infty} dx V(x) e^{iqx} \quad (S2)$$

This integral can be done analytically, leading to eq.(4).

*- Charge susceptibility equations (5) and (7).*
As discussed in the main text, we assume that the tunneling electrons can only propagate along the $x$ direction, while the wavefunction is fixed along the $y$ direction. The charge is homogeneous in this direction, $0 \leq y \leq L_y$, and $\rho(y) = 1/L_y$. Then, if we calculate the Fourier transform with respect to the $x$ coordinate, we find

$$\chi_{1D}(q_x, \omega) = \frac{1}{2\pi} \int_0^{L_y} dy \int_0^{L_y} dy' \int_{-\infty}^{\infty} dq_y \, e^{iq_y y} \chi_{2D}(q_x, q_y, \omega) \approx$$
$$\approx L_y \chi_{2D}(q_x, q_y = 0, \omega) \quad (S3)$$

- *Effective potential in the presence of a magnetic field.*
The existence of a Landau level at zero energy requires the modification of the analysis in the main text.
We define the Hamiltonian as

$$H \equiv \begin{cases} \begin{pmatrix} 0 & v_F(k_x - ik_y) \\ v_F(k_x + ik_y) & 0 \end{pmatrix} & |x| > \frac{L_x}{2} \\ \begin{pmatrix} 0 & iv_F\left(\partial_x - k_y + \frac{x}{\ell_B^2}\right) \\ iv_F\left(\partial_x + k_y - \frac{x}{\ell_B^2}\right) & 0 \end{pmatrix} & |x| < \frac{L_x}{2} \end{cases} \quad \text{(S4)}$$

where $\ell_B$ is the magnetic length. The wavefunctions are

$$\Psi_{k_y}(x) \equiv \begin{cases} \begin{pmatrix} 1 \\ e^{i\theta} \end{pmatrix} + R_{k_y}\begin{pmatrix} 1 \\ e^{-i\theta} \end{pmatrix} & x < \frac{L_x}{2} \\ A_{k_y}\begin{pmatrix} e^{(x-k_y\ell_B^2)^2/(2\ell_B^2)} \\ 0 \end{pmatrix} + B_{k_y}\begin{pmatrix} 0 \\ e^{-(x-k_y\ell_B^2)^2/(2\ell_B^2)} \end{pmatrix} & |x| < \frac{L_x}{2} \\ T_{k_y}\begin{pmatrix} 1 \\ e^{i\theta} \end{pmatrix} & x > \frac{L_x}{2} \end{cases}$$
(S5)

where $R_{k_y}$ and $T_{k_y}$ are the reflection and transmission coefficients, $A_{k_y}$ and $B_{k_y}$ are numerical coefficients, and $e^{i\theta} = (k_x + ik_y)/|\vec{k}|$.

After some algebra, we find

$$T_{k_y} = \frac{2i\sin(\theta)}{e^{i\theta}e^{k_y L_x} - e^{-i\theta}e^{-k_y L_x}} \quad \text{(S6)}$$

This expression is similar to that found in the pseudodiffusive regime. The transmission decays quickly for momenta such that $|k_y| \gg L_x^{-1}$. The wavefunction in the barrier region is

$$\Psi_{k_y}(x) = T_{k_y}\begin{pmatrix} e^{[x^2 - 2k_y\ell_B^2(x-L_x/2) - L_x^2/2]/(2\ell_B^2)} \\ e^{i\theta}e^{-[x^2 - 2k_y\ell_B^2(x-L_x/2) - L_x^2/2]/(2\ell_B^2)} \end{pmatrix} \quad \text{(S7)}$$

The correction to the effective action, using perturbation theory, is of the form

$$\delta S \approx \int d\tau \int d\tau' \left|\Psi_{k_y}[x(\tau)]\right|^2 \left|\Psi_{k_y}[x(\tau')]\right|^2 W[x(\tau) - x(\tau'), \tau - \tau'] \quad \text{(S8)}$$

We assume that $|k_y| \approx L_x^{-1}$. The leading term in $\left|\Psi_{k_y}[x(\tau)]\right|^2$ comes from the lower part of the spinor in eq.(S7) (the upper part is the sign of the magnetic field is reversed). The expression for the effective action acquires an exponential factor, $\sim e^{-[x(\tau)-x(\tau')]^2/\ell_B^2}$. The multiplication of the effective interaction by a Gaussian is equivalent to averaging the effective interaction in momentum space over a range of momenta $q \approx \ell_B^{-1}$.

*- Interactions and tunneling between localized states.*
Vacancies and other strong scatterers introduce resonances near the Dirac energy in graphene. The resulting conductivity is of order $e^2/\hbar$. Transport is due t tunneling over lengths of order of the the distance between defects, $d \approx n_d^{-1/2}$, where $n_d$ is the density of defects. As discussed in the main text, the conductivity acquires a factor $e^{-\delta S}$. The main qualitative change is that the distance $d$ replaces $L_x$ and $L_y$ in eqs. (6) and (8).